\documentclass[9pt,twocolumn,twoside]{osajnl}
\pdfoutput=1
\journal{ao} 

\setboolean{shortarticle}{false}

\ifthenelse{\boolean{shortarticle}}{\colorlet{color2}{color2b}}{\colorlet{color2}{color2}} 

\title{Automated alignment of a reconfigurable optical system using focal-plane sensing and Kalman filtering}

\author[1,*]{Joyce Fang}
\author[1]{Dmitry Savransky}

\affil[1]{Sibley School of Mechanical and Aerospace Engineering, Cornell University, 105 Upson Hall, Ithaca, NY, 14853}

\affil[*]{Corresponding author: jf648@cornell.edu}

\dates{Compiled \today}

\ociscodes{(220.1140) Alignment; (010.7350) Wave-front sensing; (120.0120) Instrumentation, measurement, and metrology; (150.5758) Robotic and machine control; (100.3010) Image reconstruction techniques.}

\doi{\url{http://dx.doi.org/10.1364/ao.XX.XXXXXX}}

\begin{abstract}
Automation of alignment tasks can provide improved efficiency and greatly increase the flexibility of an optical system. Current optical systems with automated alignment capabilities are typically designed to include a dedicated wavefront sensor. Here, we demonstrate a self-aligning method for a reconfigurable system using only focal plane images. We define a two lens optical system with eight degrees of freedom. Images are simulated given misalignment parameters using ZEMAX software. We perform a principal component analysis (PCA) on the simulated dataset to obtain Karhunen-Lo\`eve (KL) modes, which form the basis set whose weights are the system measurements. A model function which maps the state to the measurement is learned using nonlinear least squares fitting and serves as the measurement function for the nonlinear estimator (Extended and Unscented Kalman filters) used to calculate control inputs to align the system. We present and discuss both simulated and experimental results of the full system in operation.
\end{abstract}

\setboolean{displaycopyright}{true}
\usepackage[labelsep=colon]{caption}
\usepackage{subcaption}

\begin{document}

\maketitle
\thispagestyle{fancy}

\ifthenelse{\boolean{shortarticle}}{\ifthenelse{\boolean{singlecolumn}}{\abscontentformatted}{\abscontent}}{}

\section{Introduction}
Advanced optical systems are widely used in today's technology, including observing and tracing biological and chemical compounds with microscopes, creating three-dimensional scenes with virtual reality (VR) displays \cite{rolland2005head}, detecting and imaging exoplanets and disks with ground-based and space telescopes \cite{dantowitz2000ground}, and sensing and correcting wavefront aberration for medical purposes \cite{maeda2002wavefront}. An automated optical alignment system can save the time and energy spent on manually alignment. This makes the assembly process of many optical devices, including microscopes, medical sensing devices, and camera systems, more efficient. Self-aligning techniques can also improve the alignment between lenses of a virtual reality headset and human eyes. Most current VR headset models only allow a manual adjustment of the interpupillary distance (IPD). Moreover, automated alignment is very important for space optical systems. Many satellites and space telescopes cannot be serviced after their launch. A slight inaccuracy in the engineering design or disturbance during launch or on orbit can easily cause optical misalignment \cite{hartig2003orbit}. The importance and benefits of automatically aligning an optical system increase with the complexity and flexibility of the instruments themselves. Of particular interest is the ability for complex instruments to automatically align using existing internal imaging sensors, without requiring the addition of dedicated wavefront sensors, or other large changes to their basic beam paths.

Many static components in optical systems (such as reimaging and collimating optics) are bolted down after begin carefully aligned the first time. In these cases, the manual alignment procedures are time consuming and optical misalignment caused by environmental disturbances cannot be fixed. A reconfigurable system, which has multiple filters or other components in pupil and focal planes, needs the ability to self-align, and may be made more flexible if internal components are allowed to move. For example, the Gemini Planet Imager (GPI) \cite{macintosh2014first}, a ground-based instrument which includes a coronagraph and an extreme adaptive optics system for direct imaging of extrasolar planets, has automated alignment features on coronagraph components using computer vision algorithms \cite{savransky2013computer}. The closed loop control process allows GPI to achieve high precision alignment in the presence of a continuously changing gravity gradient and thermal flexure. A distributed optical system, such as an optical communication system, needs to be accurately aligned within limited space and setup time. Finally, there are cases where allowing for motion degrees of freedom creates new sensing capabilities as in interferometric devices and self-coherent imaging systems \cite{galicher2010self}.

The most widely used alignment methods relate misalignment parameters to optical wavefront error as measured by various wavefront sensing devices. One of these methods involves mapping misalignments to Zernike terms using a sensitivity table \cite{figoski1989computer,gao2004computer}. Sensitivity tables, however, are limited in their accuracy when the misalignments are large and the nonlinearity increases. Merit function regression solves this problem and is presented in Kim et al. \cite{kim2007merit}. This method estimates the misalignment by performing damped least square optimization with merit function values defined as the difference between the measured and ideal Zernike coefficients of the optical system wavefronts. Lee et al. \cite{lee2007computer} proposed a differential wavefront sampling (DWS) method for the efficient alignment of optical systems. By perturbing optical elements this technique generates a set of linear equations used to solve for the misalignment of a system. Oh et al. \cite{oh2010integration} integrated revised DWS sampling method with MFR non-linear optimization on a three-mirror anastigmat optical system. The integrated alignment method results in better alignment accuracies than standard MFR and DWS methods. Instead of using a numerical approach, Gu et al. \cite{gu2015alignment} presented a method for aligning a three-mirror anastigmatic telescope using nodal aberration theory. These methods all require measuring the wavefront error of the system. When detailed knowledge of a wavefront is required for alignment, optical systems are designed to include a dedicated wavefront sensor, such as Shack-Hartmann wavefront sensor.  This increases the complexity of the system, and, more importantly, can introduce non-common path errors. 

There exist, however, various wavefront sensing schemes employing the primary system sensor and eliminating the need for dedicated wavefront sensors \cite{pueyo2009optimal,groff2013focal}. These techniques are already being applied to current scientific instrumentation \cite{savransky2012focal}. Some focal-plane wavefront sensing methods use pupil plane masking or multiple detectors to introduce phase diversity and reconstruct wavefront error \cite{gonsalves1979wavefront,wang2015zernike,martinache2013asymmetric}. The image moment-based wavefront sensing (IWFS) method \cite{lee2012fine} uses image moment of measured point spread function (PSF) for alignment correction. Focus diversity (FD) is introduced to break the nonlinearity and allows the system to sense full-field wavefront aberration. These methods can greatly improve wavefront sensing and optical alignment techniques, and can be applied to automated optical alignment.

We propose a method which corrects the misalignment of an optical systems with existing internal imaging instruments in the system, such as a focal plane camera, there by saving the extra resources and space need for splitting the beam, and avoiding throughput loss and non-common path error. Moreover, this approach makes it easier to retrofit an existing optical system to perform self-alignment since the major difference is changing the static optical components to kinematic ones. We present a new sensing and control method for aligning a reconfigurable optical system. We demonstrate the ability to align a two lens system using only a focal plane camera. An optical model of a monochromatic beam, two moving lenses, and a science camera is connected to a closed-loop control system. We implement an iterated extended Kalman filter (IEKF) and unscented Kalman filter (UKF) to estimate the states in the control process. Our current alignment methodology is focused on narrow field of view (FOV) systems and focuses on the on-axis signal.  However, the basic approach can be extended to also consider off-axis sources and be made relevant for systems with larger FOVs. Examples of small FOV systems currently in use include light detection and ranging (LiDAR) systems for detail local mapping \cite{wandinger2005introduction}, high contrast imaging system for imaging exoplanets near bright stars \cite{guyon2006theoretical}, and high resolution satellites \cite{dial2003ikonos,toutin2002quickbird}. In Section \ref{sec:model_methods}, we define the optical model and control scheme, and discuss the methods used in modeling and estimation, including Karhunen-Lo\`eve modal reconstruction, model fitting, and state estimation. We present the simulation result and introduce the experimental setup. In Section \ref{sec:results}, the experimental result is presented. Both image reconstruction and closed-loop state estimation are shown. In Section \ref{sec:discussion}, we discuss the result of our current system, and lay out our next steps.

\section{Model and Methodology}\label{sec:model_methods}

In this section we define the optical and control model, discuss image processing methods, model fitting, closed-loop control system, and experimental setup.

\subsection{Model}\label{sec:model}

\begin{figure}[htbp]
\centering
\includegraphics[trim=15mm 120mm 10mm 0mm,clip=true,width=\linewidth]{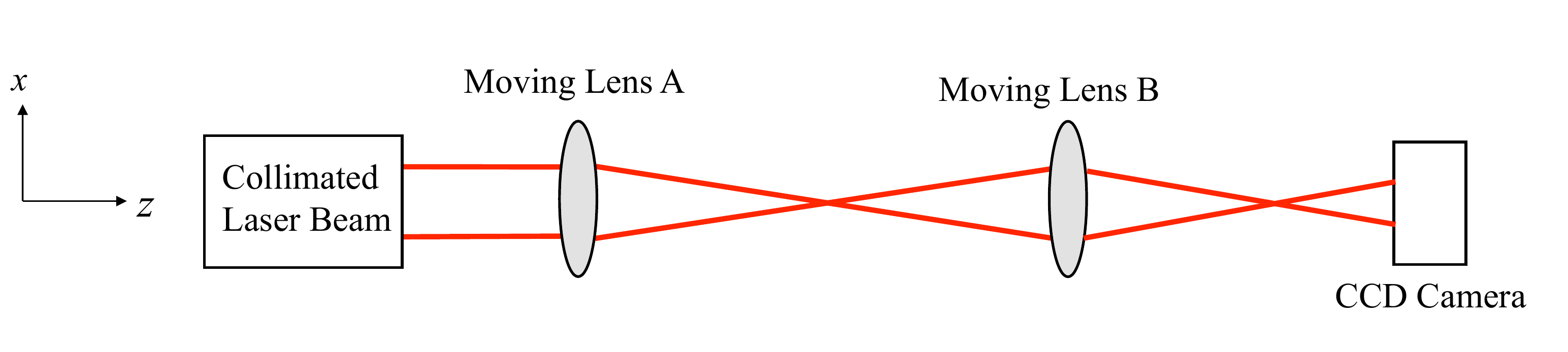}
\caption{Two lens optical system. A collimated Gaussian beam is passed through two moving lenses A and B, and focuses on a CCD camera.}
\label{fig:two_lenses_model}
\end{figure}

Figure \ref{fig:two_lenses_model} shows a two lens optical system. The two moving lenses are represented as gray ellipses, the collimated laser beam is represented by the red line along the $z$-axis, the $x$-$y$ plane is normal to the beam path, and the $x$-axis is along the vertical. The collimated Gaussian beam passes through two moving lenses A and B, and is focused on a CCD camera. The focal lengths of lens A and B are set as $200$ and $100$ $mm$, respectively, the image plane has pixel size $4.54$ $\mu m$, and the laser beam has wavelength 635 $nm$. The distance between the collimated laser beam and lens A is 50 $mm$, and the distance between lens A and lens B is 400 $mm$. The CCD camera is placed at a distance of 212 $mm$ after lens B. Our goal is to calibrate the moving lenses, which have a total of 8 degrees of freedom - shift in $x$ and $y$ direction, tip and tilt for each lens. This is a simple model where the 
despace misalignments of both lens A and B are assumed to have smaller influence on the system compared with lateral motions, and the shift along the $z$ axis of both lenses are not included in the model. A column vector $\mathbf{x} = [S_{xA}, S_{yA}, S_{xB}, S_{yB}, T_{xA}, T_{yA}, T_{xB}, T_{yB}]^T$ is used to describe the state of our system.

\begin{figure}[htbp]
\centering
\includegraphics[trim=0mm 0mm 0mm 0mm,clip=true,width=\linewidth]{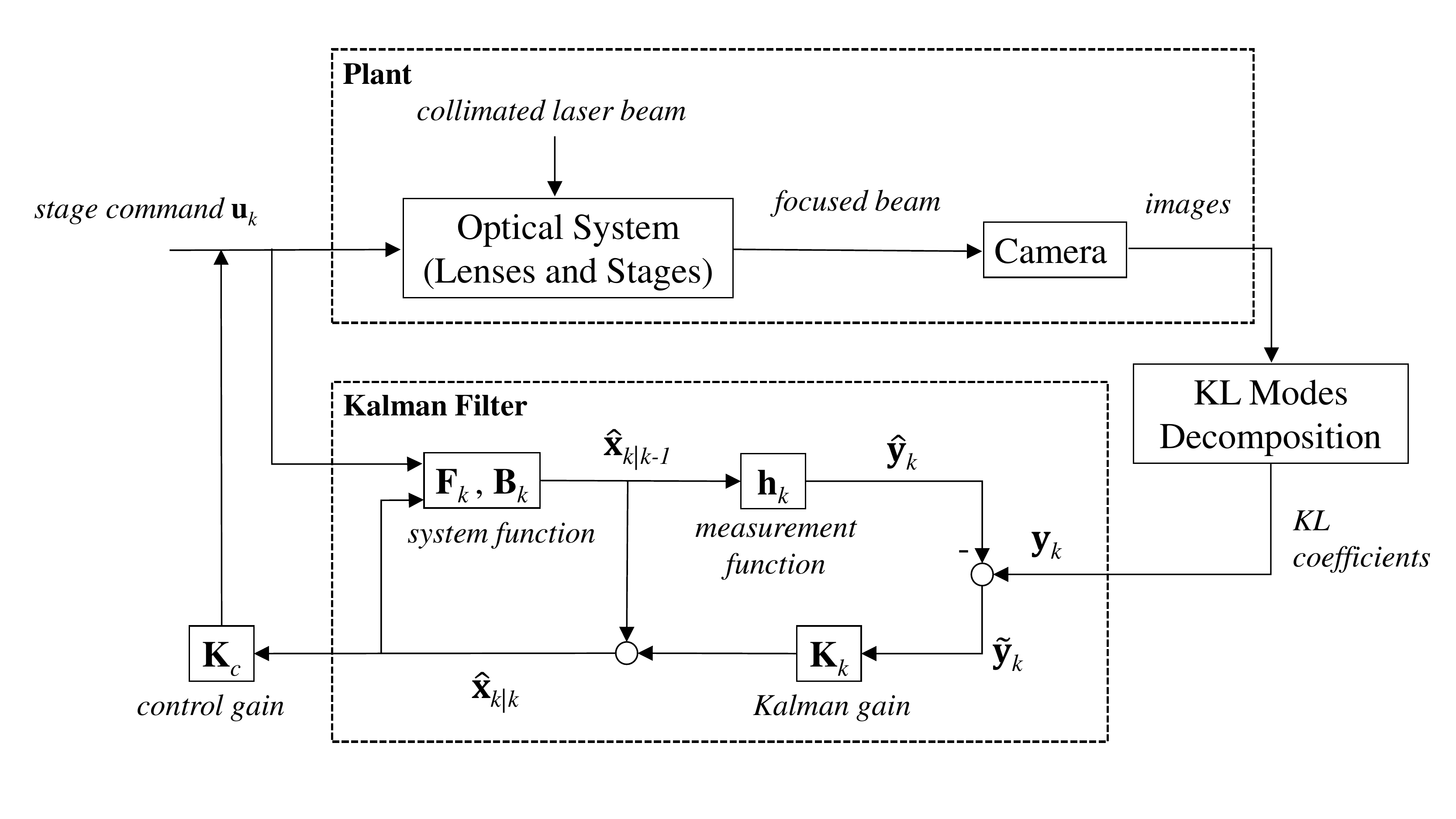}
\caption{Schematic of control system. Upper dashed block represents the optical system in Figure \protect\ref{fig:two_lenses_model}, and lower dashed block represents a Kalman filter.}
\label{fig:control_scheme}
\end{figure}

Figure \ref{fig:control_scheme} shows a schematic of the self-algining control system. The optical model in Figure \ref{fig:two_lenses_model} is in the upper dashed block (Plant), and the lower dashed block represents a Kalman filter in closed-loop control system. The images captured from the camera are projected onto Karhunen-Lo\`eve (KL) modes obtained from principal component analysis (PCA), which will be discussed in the next section, with the corresponding KL weights serving as the measurements of the control system. The measurements are sent to the Kalman filter to compare with the measurement predicted from our measurement model function (section \ref{subsec:meas_fun}). The state estimate predicted by the Kalman filter is fed back to correct the misalignment (section \ref{subsec:kf}).

\subsection{Image Processing}\label{subsec:img_processing}
We simulate images given misalignment parameters using ZEMAX software. The prescription in ZEMAX is set as the optical system described in section \ref{sec:model}. Thorlabs lens LB1945 and LB1676 are imported as lens A and lens B, respectively. The laser, lenses, and camera parameters are chosen to model the conditions in the experiment as shown in Table \ref{tab:exp_spec}.
Misalignment of the lenses introduces wavefront aberrations into the optical system, resulting in  motion and shape changes to the nominally axisymmetric Gaussian spot in focal plane images. Figure \ref{fig:sim_img} shows a sample misalgined image in $250 \times 250$ pixels. Our first image processing step is to fit a 2D Gaussian to the image to obtain the center position of the Gaussian spot of the image, and then perform PCA to decompose the image dataset into KL modes. 

\subsubsection{Gaussian Fitting and COM}
\begin{figure}[htbp]
\centering
        \includegraphics[trim=0mm 18mm 0mm 10mm,width=\linewidth]{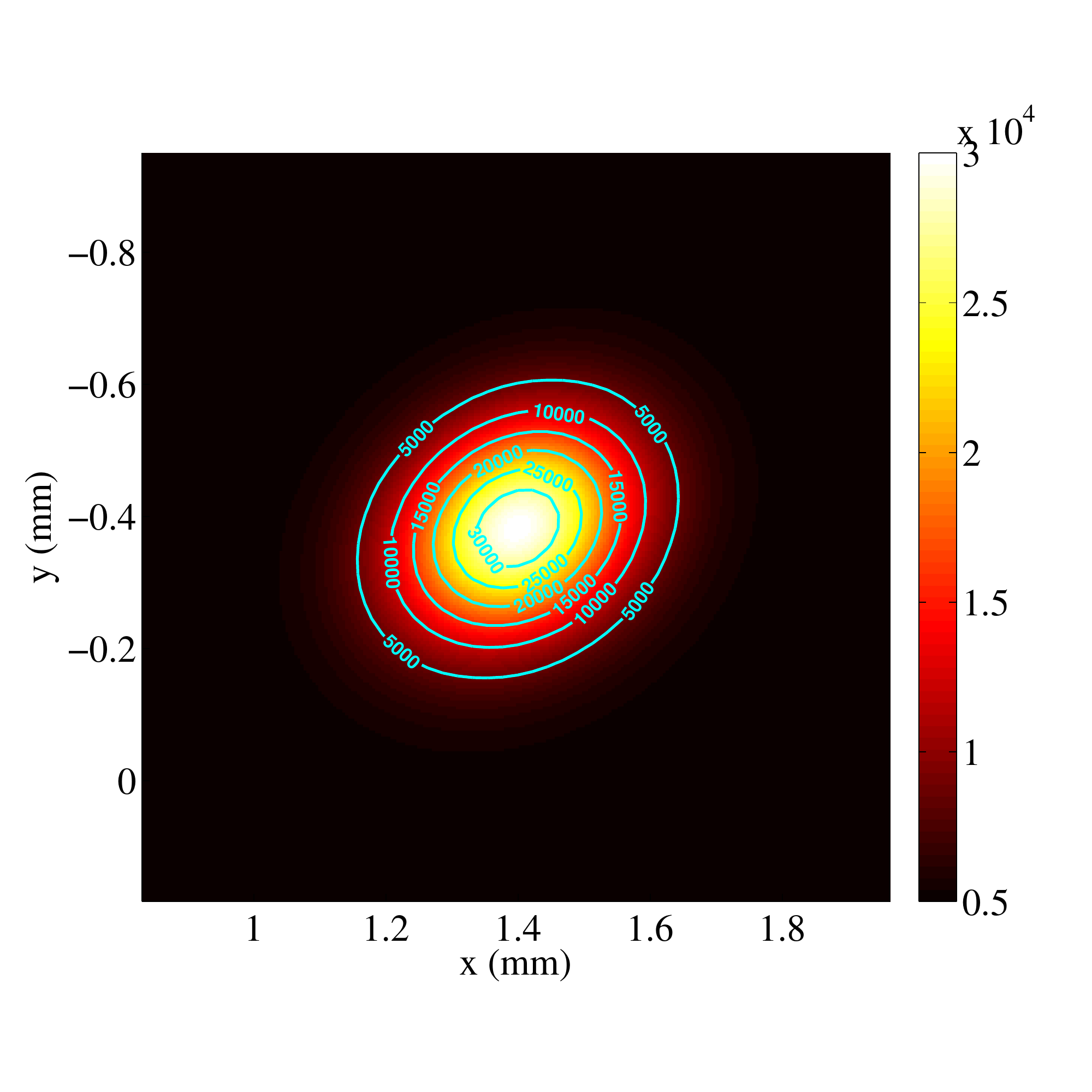}
        \caption{Subframe of the simulated image from Zemax and its 2D Gaussian fitting in contour plot. Gaussian center $C_x$ and $C_y$ will be used as our measurements.}
        \label{fig:sim_img}
\end{figure}

We apply a Gaussian fit to the subframe:
\begin{equation}
\label{eq:gaussian}
\begin{split}
T(x,y)&=G_1 + G_2\exp\left(\frac{-(\frac{x'}{a})^2 - (\frac{y'}{b})^2}{2}\right)\\
x'& = (x-C_x)\cos\phi - (y-C_y)\sin\phi \\
y' &= (x-C_x)\sin\phi + (y-C_y)\cos\phi
\end{split}
\end{equation}
The unknown parameters ($C_x$, $C_y$) represent the center position of the Gaussian spot, ($G_1$, $G_2$) represent the Gaussian coefficients, ($a$, $b$) represent the semi-major and minor axis, and $\phi$ represents the rotational angle of the ellipse. Figure \ref{fig:sim_img} shows the Gaussian fitting of the simulated image in contour plot. The Gaussian center $C_x$ and $C_y$ can be obtained and will be used as our measurements.
\subsubsection{PCA and Karhunen-Lo\`eve Modes}\label{sec:pca}
We perform a PCA using the Karhunen-Lo\`eve Transform (KLT) to create an orthogonal basis of eigenimages \cite{rao2000transform,soummer2012detection}. KLT method decomposes observed signals into a combination of linearly independent modes called principal components. The observed signals are the image data set we collected from ZEMAX. In this section we call the linearly independent modes KL modes.

We collect the image dataset by scanning through eight state variables $S_{xA}$, $S_{yA}$, $S_{xB}$, $S_{yB}$, $T_{xA}$, $T_{yA}$, $T_{xB}$, and $T_{yB}$. Each state is perturbed with 3 misaligned values $-\delta$, 0, and $\delta$, where $\delta$ is a small misalignment for each state. We perturb the shift and tip-tilt by 0.4 $mm$ and 4 $degree$ respectively. The collected images include all combinations of the perturbation on the states. This results in a total of $3^8$ scanned images. We capture a fixed size subframe ($N_p \times N_p$) around the Gaussian center $C_x$ and $C_y$. The subframe image matrix is reshaped into a $p$-element column vector $\mathbf{v}_i$, where $p = N_p^2$ and $i$ indicates the image number. We use the vector-mean-subtracted value of the image vector
   \begin{equation}
	\label{eq:mean}
	\mathbf{\bar{v}}_i = \mathbf{v}_i - \mu(\mathbf{v}_i) \,,
   \end{equation}
where $\mu(\cdot)$ is a mean operator. A large matrix containing all the scanned data can be obtained as
   \begin{equation}
	\label{eq:R_bar}
	\mathbf{\bar{V}} = [\bar{\mathbf{v}}_1, \bar{\mathbf{v}}_2, ..., \bar{\mathbf{v}}_n] \,.
   \end{equation}
The image-to-image covariance matrix of the scanned data set is given by
   \begin{equation}
	\label{eq:S}
	\mathbf{S} = \frac{1}{p-1}\mathbf{\bar{V}}^T \mathbf{\bar{V}} \,,
   \end{equation}
where $\mathbf{S}$ is an $n\times n$ matrix. We perform an eigendecomposition of covariance $\mathbf{S}$ and obtain matrices $\mathbf{\Phi}$ and $\mathbf{\Lambda}$ such that 
   \begin{equation}
	\label{eq:eigen}
	\mathbf{S}\mathbf{\Phi} = \mathbf{\Phi}\mathbf{\Lambda} \,,
   \end{equation}
   where $\mathbf{\Phi}$ is an $n\times n$ matrix whose columns are the eigenvectors of $\mathbf{S}$, and $\mathbf{\Lambda}$ is an $n\times n$ diagonal matrix whose entries are the corresponding eigenvalues.
   The KL transform matrix is then
   \begin{equation}
	\label{eq:Z}
	\mathbf{Z} = \bar{\mathbf{V}}\mathbf{\Phi} \,,
   \end{equation}
where $\mathbf{Z}$ is a $p \times n$ matrix whose columns are the KL modes. 

\begin{figure}[htbp]
\centering
\includegraphics[trim=25mm 15mm 18mm 5mm,clip=true,width=\linewidth]{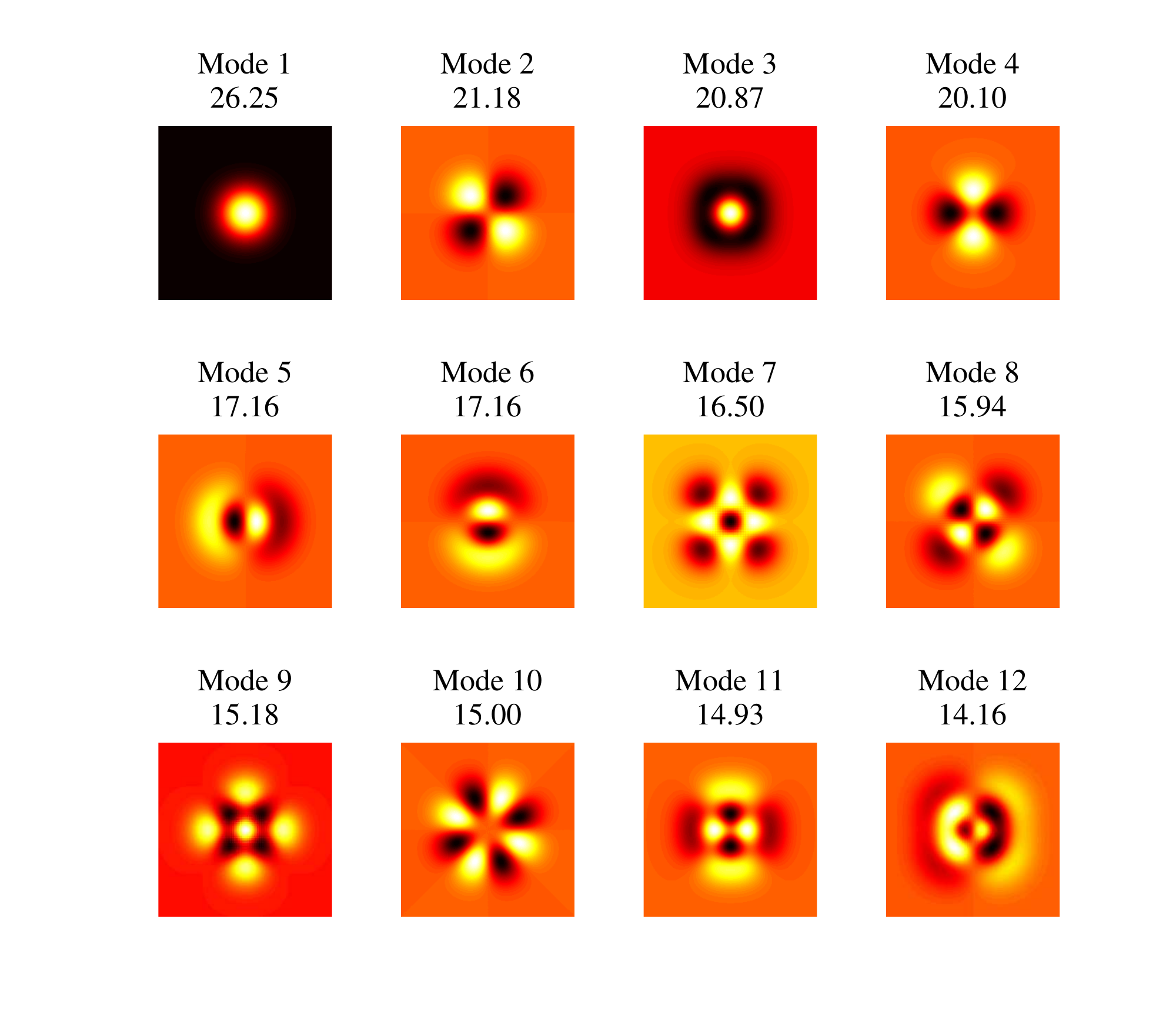}
\caption{First 12 KL modes obtained from PCA decomposition with subframe $250 \times 250$ pixels. Each image is plotted under different intensity scale and its corresponding eigenvalue is shown under the mode number in log scale.}
\label{fig:KLmodes}
\end{figure}

Figure \ref{fig:KLmodes} shows the first 12 KL modes (in order of decreasing eigenvalue) and the corresponding eigenvalue in log scale. Each image has a frame size of $250 \times 250$ pixels, and all images are independently stretched to show details of the mode shapes.

\begin{figure}[htbp]
\centering
        \begin{subfigure}[t]{0.45\linewidth}
                \includegraphics[trim=0mm 0mm 0mm 12mm,clip=true, width=\linewidth]{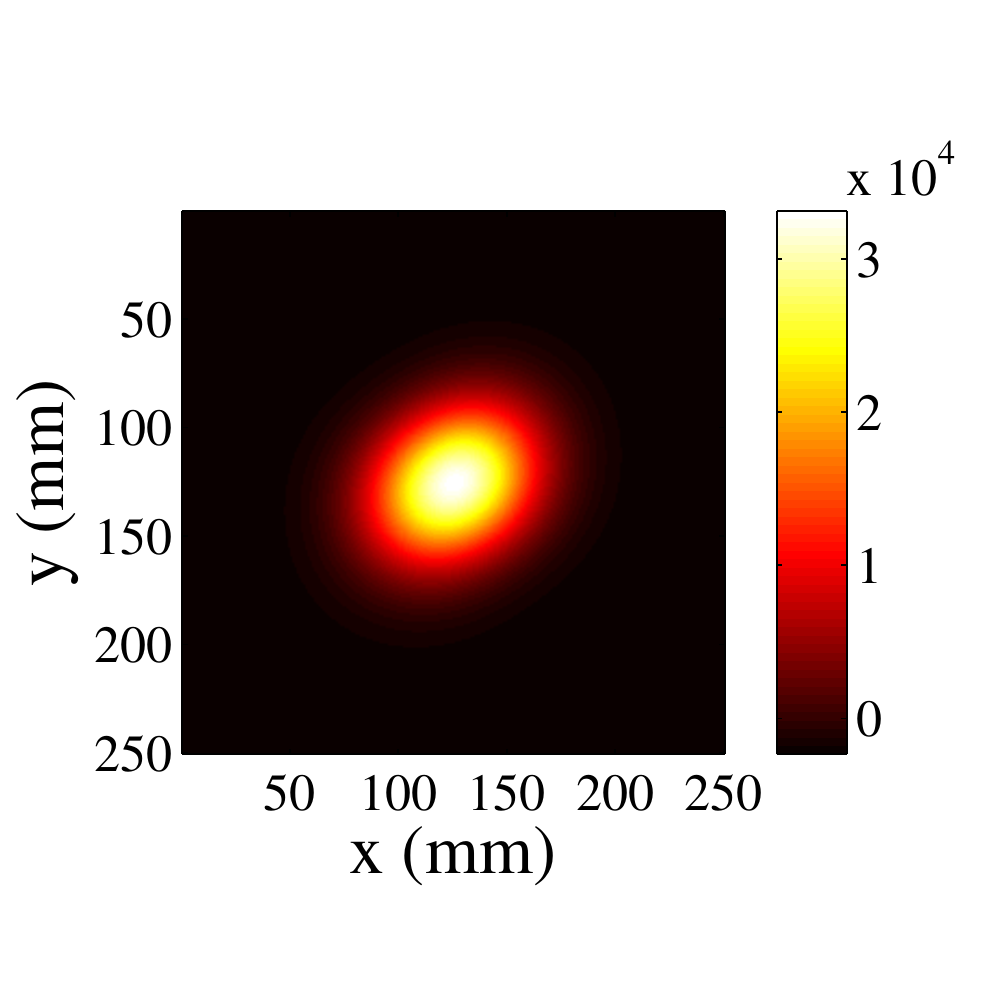}
                \caption{Reconstructed image}
                \label{subfig:recon_img}
        \end{subfigure}\hspace{1ex}
        \begin{subfigure}[t]{0.45\linewidth}
                \includegraphics[trim=0mm 0mm 0mm 15mm,clip=true, width=\linewidth]{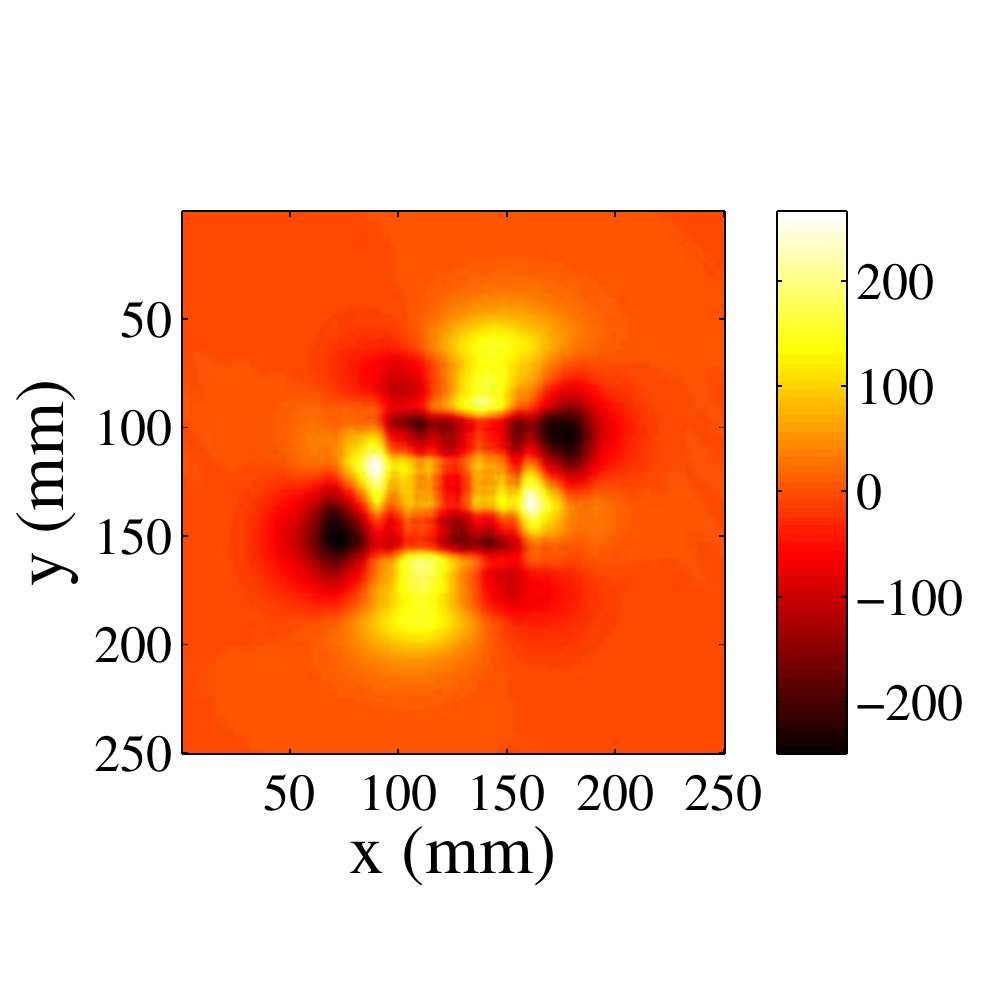}
                \caption{Subtracted image}
                \label{subfig:substract_img}
        \end{subfigure}
        \caption{Reconstruction of simulated image in Figure \protect\ref{fig:sim_img} using the first six KL modes. Subtracted image shows the difference ($\mathbf{\bar{v}}_i - \mathbf{\bar{c}}_i$) between the simulated image in Figure \protect\ref{fig:sim_img} and reconstructed images.}
        \label{fig:recons}
\end{figure}

The image $\mathbf{\bar{v}}_i$ can be reconstructed as a weighted sum of the first $m$ KL modes with coefficients
   \begin{equation}
	\label{eq:w2}
	\mathbf{w}_i = \mathbf{Z}_m^{\dagger} \mathbf{\bar{v}}_i \,,
   \end{equation}
where $(\cdot)^{\dagger}$ is the pseudoinverse of a matrix, and $\mathbf{Z}_m$ contains the first $m$ KL modes. The reconstructed image $\mathbf{\bar{c}}_i$ can be calculated as
   \begin{equation}
	\label{eq:recon}
	\mathbf{\bar{c}}_i = \mathbf{Z}_m\mathbf{w}_i \,,
   \end{equation}
   where $\mathbf{w}_i$ is an $m$-element column vector of the coefficients calculated above. Figure \ref{subfig:recon_img} shows the reconstructed image $\mathbf{\bar{c}}_i$ of the simulated image in Figure \ref{fig:sim_img} using the first six KL modes. The subtracted image in Figure \ref{subfig:substract_img} shows the difference ($\mathbf{\bar{v}}_i - \mathbf{\bar{c}}_i$) between the simulated image in Figure \ref{fig:sim_img} and the reconstructed image.
   
\begin{figure}[htbp]
\centering
\includegraphics[trim=30mm 15mm 15mm 0mm,clip=true,width=\linewidth]{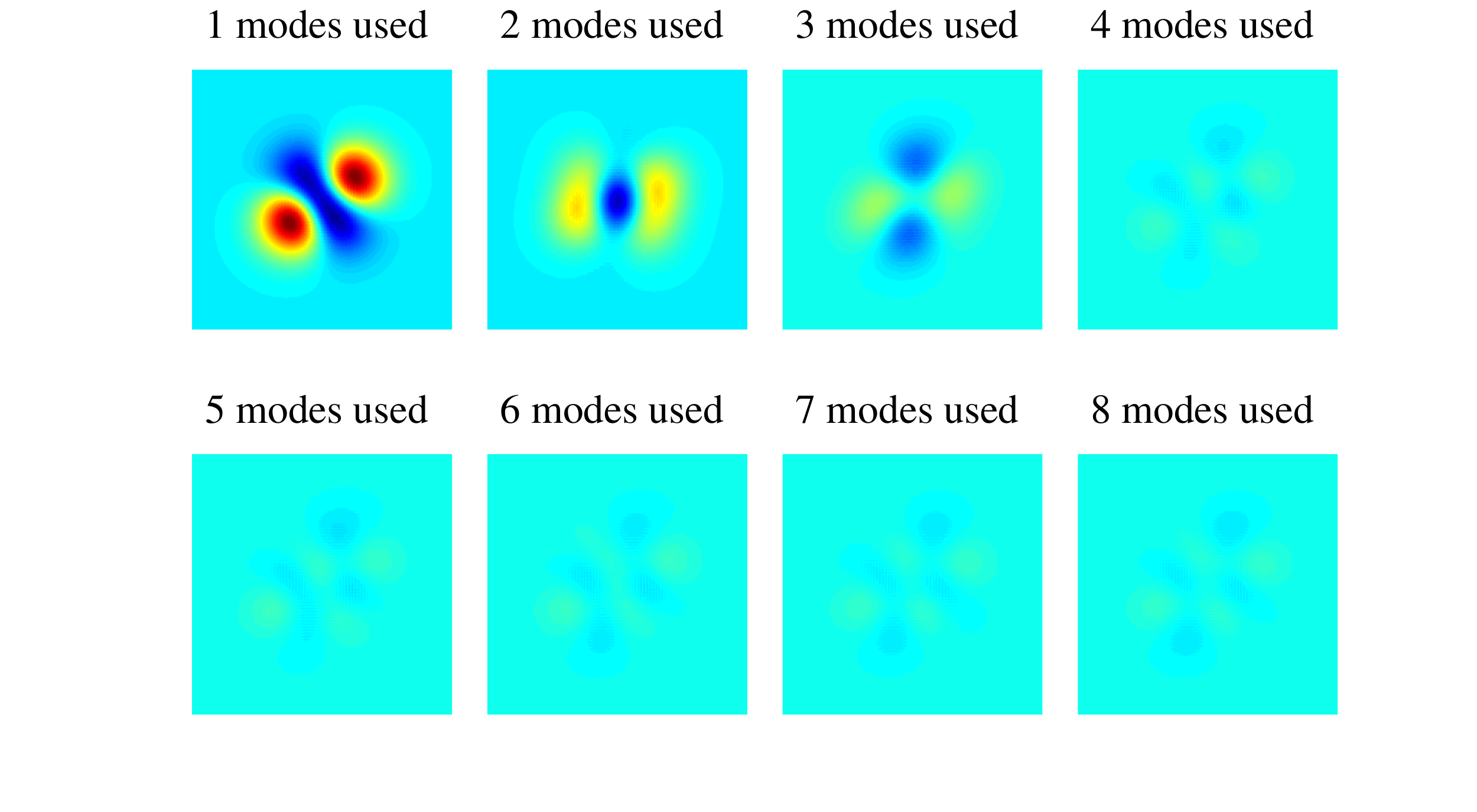}
\caption{Residual error with modes 1 - 8 used in image reconstruction. Images are plotted under the same intensity scale. The reconstruction error decreases gradually as the number of modes used increases.}
\label{fig:recon_sub}
\end{figure}

\begin{figure}[htbp]
\centering
\includegraphics[trim=0mm 0mm 0mm 0mm,clip=true,width=\linewidth]{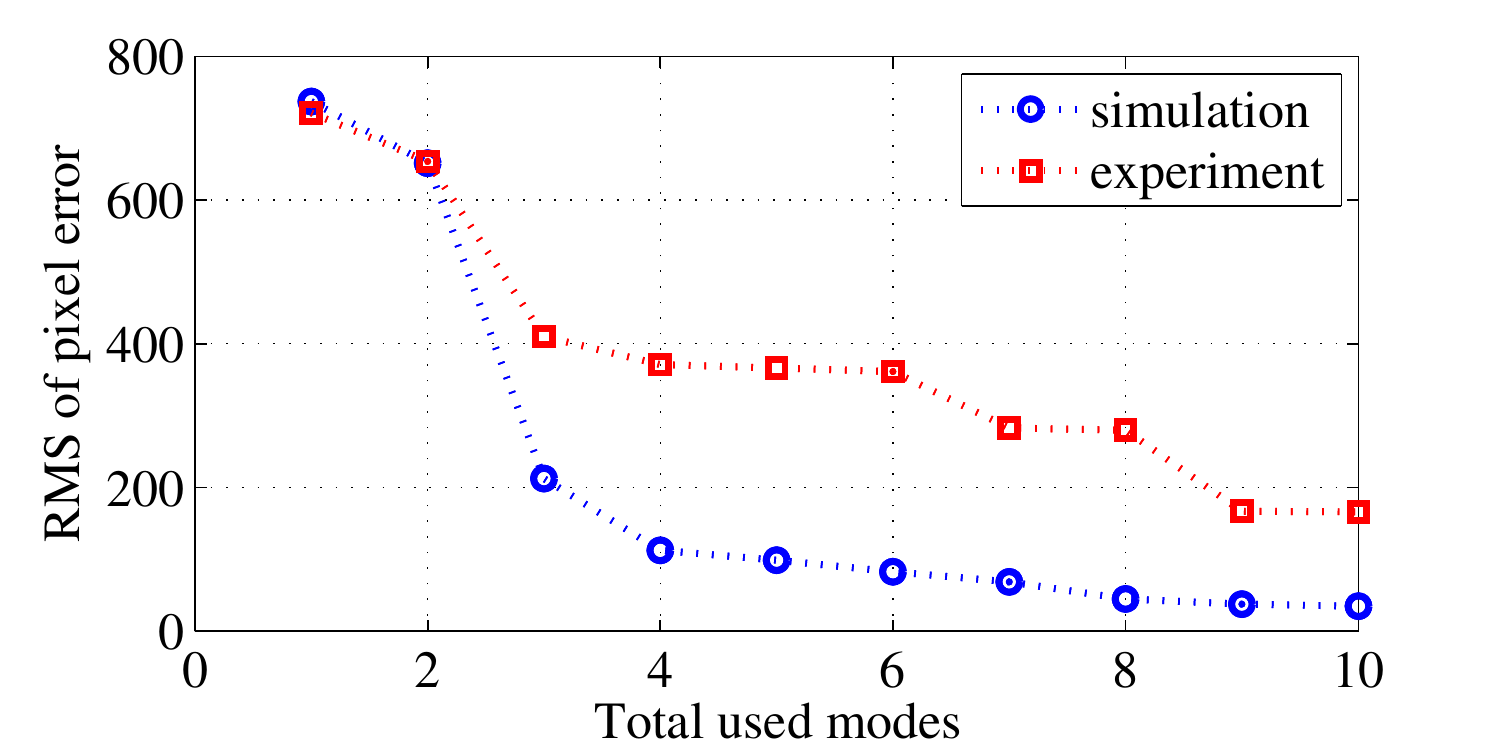}
\caption{Reconstruction RMS pixel error using the first 10 modes. The blue circle markers and red square markers represent errors in the simulation and experiment, respectively.}
\label{fig:recon_rms_error}
\end{figure}

Figure \ref{fig:recon_sub} shows the reconstruction error using the first eight modes to reconstruct the image, with all of  the images plotted on the same intensity scale. As expected, we can see that the reconstruction error decreases gradually as the number of mode used increases. The RMS pixel errors $\epsilon_i$ of the reconstruction of image $i$ is defined as
   \begin{equation}
	\label{eq:rms_pixel_error}
	\epsilon_i = \frac{\sqrt{(\mathbf{\bar{v}}_i - \mathbf{\bar{c}}_i)^T(\mathbf{\bar{v}}_i - \mathbf{\bar{c}}_i)}}{p},
   \end{equation}
and is used as a metric for the quality of the image reconstruction. The blue circle markers in Figure \ref{fig:recon_rms_error} shows the RMS error $\epsilon_i$ after the frist 10 modes used. The RMS pixel error shown is the average over all of the training data.

\subsection{Measurement Function}\label{subsec:meas_fun}
The weights of KL modes 2-6, normalized by mode 1, together with the Gaussian center $C_x$ and $C_y$ are used as the measurements. The measurement $\mathbf{y}$ can be written as 
   \begin{equation}
	\label{eq:coeff}
	\mathbf{y} = \big[\frac{w_2}{w_1},\frac{w_3}{w_1},\frac{w_4}{w_1},\frac{w_5}{w_1},\frac{w_6}{w_1},C_x,C_y\big].
   \end{equation}
The simulated image set described in Section \ref{sec:pca} was obtained by scanning through the eight system states and generating 6,561 images for the KL mode decomposition. Now, we generate 60,000 images given random misaligned states to train the measurement function. Two thirds of these images (40,000) are used as our training set, and the remaining one third (20,000) as the test set. 

We perform a nonlinear least squares fitting on the training set using the Levenberg-Marquardt algorithm \cite{more1978levenberg}. The nonlinear measurement model function $\mathbf{h}$ is learned to predict measurement $\mathbf{y}$, computed as $\hat{\mathbf{y}} = \mathbf{h}(\mathbf{x})$. Each nonlinear function $h_j$ is a second order polynomial which maps the misaligned state $\mathbf{x}$ to predict measurement $\hat{\mathbf{y}}$, where $j$ is the number of measurement from 1 to 7. The coefficients of the second order polynomial is provided in Table \ref{table:h} in appendix A. The error between the simulated and predicted measurements is $\mathbf{e} = \hat{\mathbf{y}}-\mathbf{y}$, and the Normalized Root Mean Square Error (NRMSE) of element $j$ in $\mathbf{e}$ can be calculated as
\begin{equation}
	\label{eq:NRMSE}
 	\mathrm{NRMSE}(e_j) = \frac{\sqrt{(\sum_{i=1}^{n} e_{j,i})/ n}}{\max{y_j} - \min{y_j}} \,.
\end{equation}
where $i$ ranges from 1 to $n$ for $n$ points in the dataset. Table \ref{tab:error} shows the NRMSE of the prediction on measurement 1 to 7. The NRMSE are calculated to ensure the model does not suffer from overfitting.
\begin{table}
\caption{NRMSE of measurement $y_1$ to $y_7$. Both training and test error are computed to ensure the model is not overfitted.}
\vspace{-3ex}
\label{tab:error} 
\begin{center}
\resizebox{\columnwidth}{!}{
\begin{tabular}{l*{6}{c}r} \textit{NRMSE (\textperthousand)} & \textit{$y_1$} & \textit{$y_2$} & \textit{$y_3$} & \textit{$y_4$}& \textit{$y_5$} & \textit{$y_6$} & \textit{$y_7$} \\ \hline 
Training set & 3.26 & 5.51 & 4.05 & 3.97 & 4.18 & 1.05 & 1.06 \\ 
Test set & 3.27 & 5.54 & 4.10 & 4.07 & 4.27 & 1.05 & 1.06 \\
\end{tabular}
}
\end{center}
\vspace{-4ex}
\end{table}

\begin{figure}[htbp]
\centering
\includegraphics[trim=15mm 0mm 15mm 0mm,clip=true,width=\linewidth]{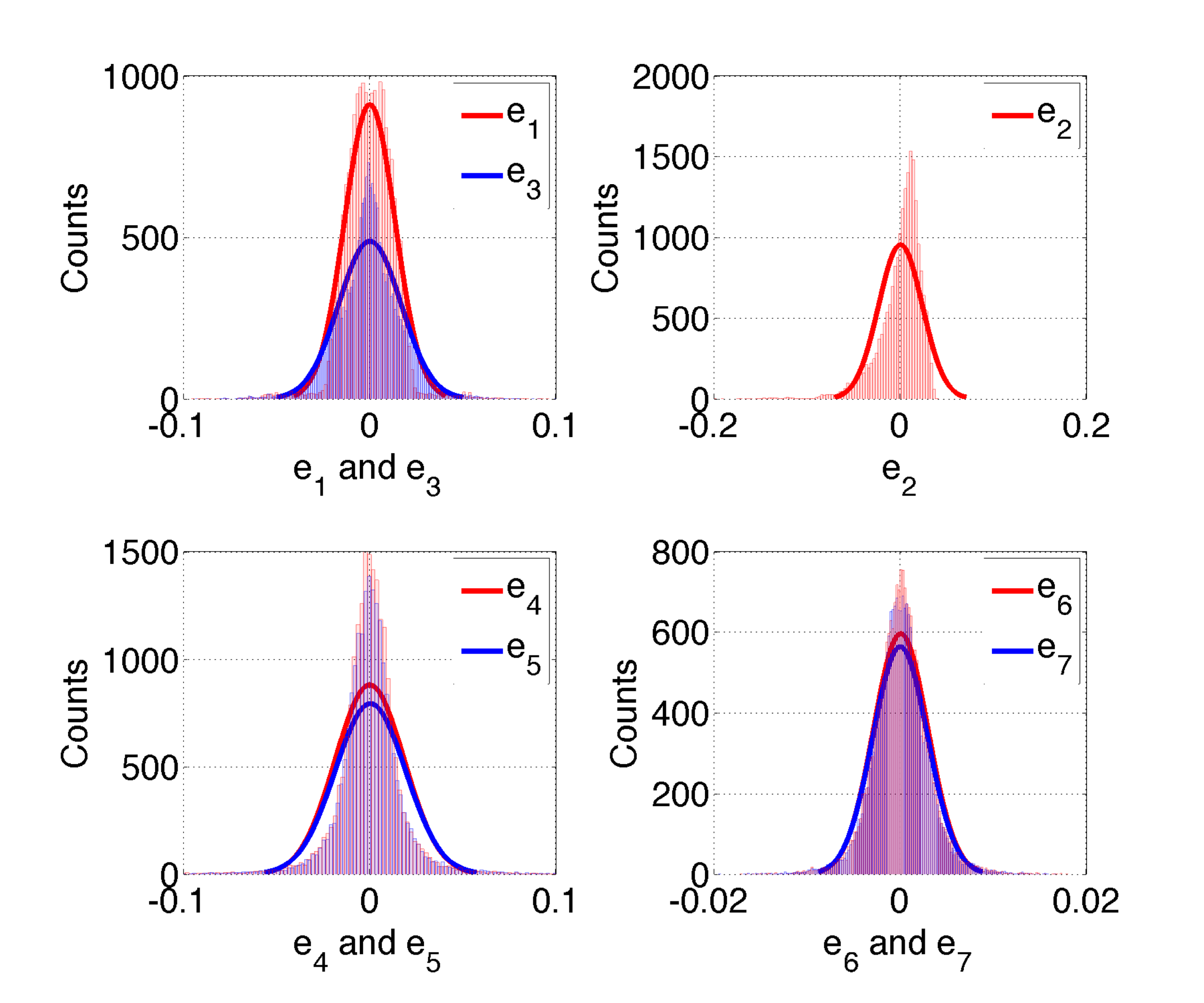}
\caption{Histogram of the residual $e_1$ to $e_7$ and their best fitted normal distribution. The closer the error distribution is to normal distribution, the more reliable the model is in Kalman filtering.}
\label{fig:train_stats}
\end{figure}

We plot the histogram of error $\mathbf{e}$ as shown in Figure \ref{fig:train_stats}. The blue and red lines are the best-fit normal distribution functions of the error distributions. The more Gaussian the distribution of the error, the better our measurement model performs in Kalman filtering. The error covariance matrix $\mathbf{R}_{model}$ is calculated as 
\begin{equation}
	\label{eq:Rmodel}
 	\mathbf{R}_{model} = \mathbf{e}\mathbf{e}^T,
\end{equation}
and used as part of the measurement covariance matrix in Kalman filtering.

\subsection{Kalman Filtering}\label{subsec:kf}
Our state space representation is 
      \begin{equation}
	\label{eq:KF}
	\begin{split}
 	\mathbf{x}_k &= \mathbf{F}_k \mathbf{x}_{k-1} + \mathbf{B}_k \mathbf{u}_k + \mathbf{q}_k \\
	\mathbf{y}_k &= \mathbf{g}_k(\mathbf{x}_k) + \mathbf{r}_k \,,
	\end{split}
   \end{equation}
where the state transition matrix $\mathbf{F}_k$ and the control input matrix $\mathbf{B}_k$ are taken to be identity matrices. The process noise $\mathbf{q}_k$ is a zero mean Gaussian with covariance $\mathbf{Q}_k$, $\mathbf{q}_k \sim N(0,\mathbf{Q}_k)$. The observation function $\mathbf{g}_k$ maps the misaligned states $\mathbf{x}_k$ to the measurement $\mathbf{y}_k$, and $\mathbf{r}_k \sim N(0, \mathbf{R}_k)$ is the measurement noise. The observation function is modeled by nonlinear measurement function $\mathbf{h}$ learned in section \ref{subsec:meas_fun}.

Kalman filtering is a sequential estimation algorithm, using a series of observations, together with statistical models of noise, to predict partially observed variables in a dynamic system. It iterates over two steps, dynamic propagation and measurement update. Kalman filtering is widely used in optical state estimation and wavefront control, including linear \cite{redding2004optical,lou2005jwst} and nonlinear \cite{riggs2015wavefront} filters. In this section, an iterated extended Kalman filter \cite{julier2004unscented} (IEKF) and an unscented Kalman filter \cite{wan2000unscented, julier1997new} (UKF) are used to estimate the misaligned states. Extended Kalman filter (EKF) is a nonlinear version of Kalman filter which approximate the mean and covariance of current estimate using local linearization of the nonlinear function. IEKF is an iterative version of EKF which insure convergency in the measurement update step \cite{stengel2012optimal,hedrick2005control}. Since our state transition ($\mathbf{F}_k$, $\mathbf{B}_k$) is linear, the nonlinear approximation only occurs in the measurement update step where we calculate the Jacobian of nonlinear function $\mathbf{h}$. UKF is also a nonlinear Kalman filter which uses unscented transform to estimate Gaussian distribution. The mean and covariance of state estimates are approximated by sigma points generated in the algorithm. The general process of Kalman filtering is shown as the lower dashed block in Figure \ref{fig:control_scheme}.

We generate process noise with covariance $\mathbf{Q}_k$ and measurement noise with covariance $\mathbf{R}_{meas}$ in the simulation. The processing noise is generated to model the stages in the experiment and has standard deviation $[0.005, 0.005, 0.005, 0.005, 0.04, 0.04, 0.04, 0.04]^T$. The measurement noise $\mathbf{R}_{meas}$ is estimated by collecting many stationary images at multiple position in the experiment, and has standard deviation $[0.0029, 0.0082, 0.0050, 0.0219, 0.0354, 0.00025, 0.00026]^T$. The total measurement noise covariance is $\mathbf{R}_k = \mathbf{R}_{model} + \mathbf{R}_{meas}$. The state estimate and estimate covariance in Kalman filtering are denoted as $\hat{\mathbf{x}}$ and $\hat{\mathbf{P}}$. We implement an IEKF and a UKF given initial guesses of the state estimate $\hat{\mathbf{x}}_0$ and state estimate covariance $\hat{\mathbf{P}}_0$.

\begin{figure}[htbp]
\centering
\includegraphics[trim=0mm 5mm 15mm 10mm,clip=true,width=\linewidth]{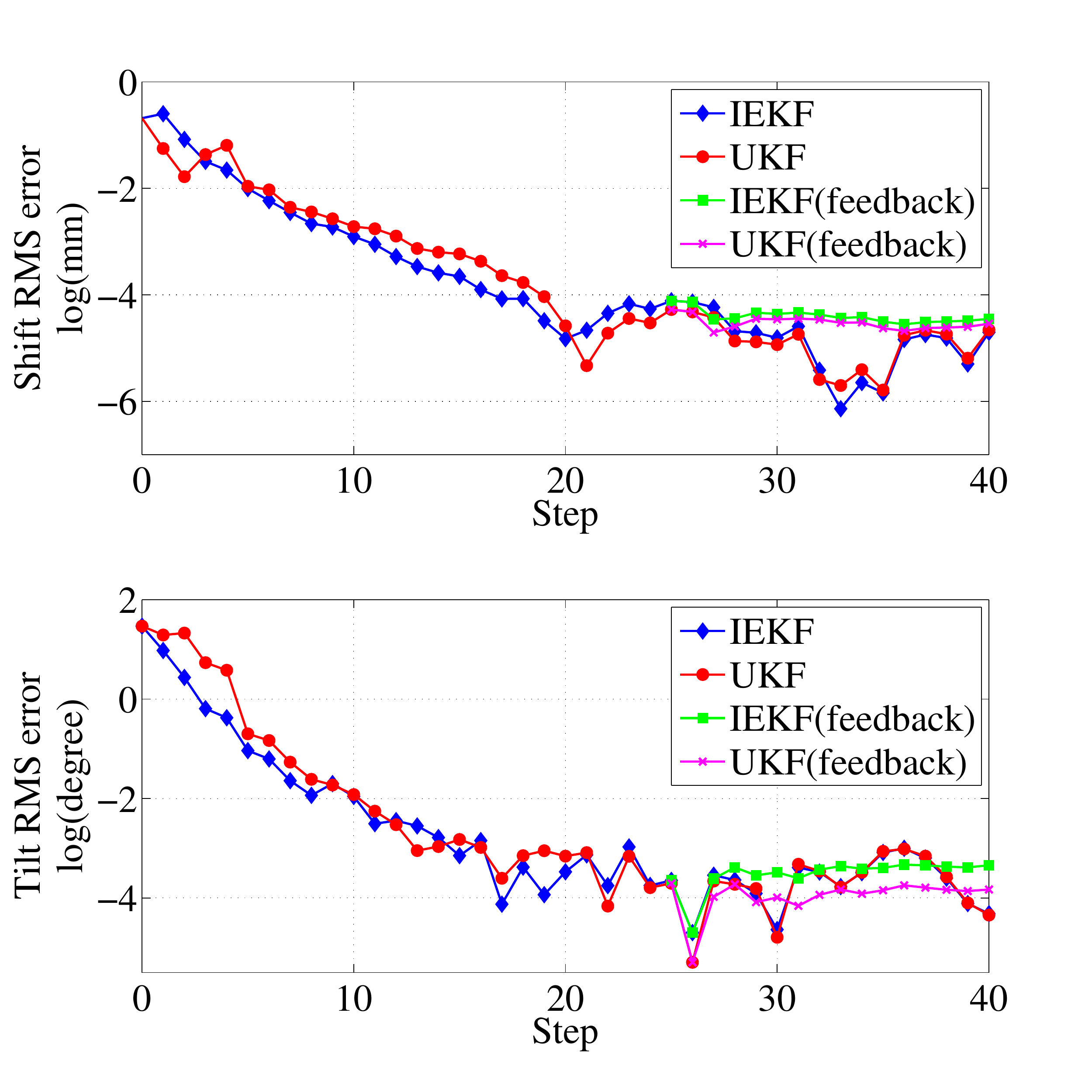}
\caption{RMS state residuals of IEFK and UKF in the simulation. Blue diamond line and red circle line represent IEKF and UKF estimation. Green square line and the magenta cross line represent the state residuals with full state feedback after the 25 step in IEKF and UKF, respectively.}
\label{fig:kf_sim}
\end{figure}

Figure \ref{fig:kf_sim} shows the RMS state residual plot of IEKF and UKF given random control input in the simulation. The lines with blue diamond red circle markers represent the IEKF and UKF estimation, respectively. Both IEKF and UKF achieve approximately $6\mu m$ error in shift and $0.02$ $degree$ error in tip and tilt. The lines with green square and magenta cross markers show the state residuals with full state feedback ($\mathbf{u}_k = -\hat{\mathbf{x}}_{k-1}$) after the 25th step in the IEKF and UKF estimation. Instead of feeding back the state estimate as the control input, we collect information by giving random inputs away from the center in the beginning of the closed-loop process. The reason we need a random walk for our system is to produce phase diversity which is not available using a single focal plane image. The perturbation needs to be above a certain value to have diversity for us to track the state, and the number of perturbations needed depends on the initial guesses of the state estimate. The step at which we start feedback is decided by the state covariance obtained in IEKF or UKF. We start full state feedback after the state estimate covariance converges and remain stable for a few steps. In the simulation we are showing one of the worst cases where the state residual decreases gradually. This indicates the initial misalignment is at a position from which our algorithm takes a long time to converges. In most of the cases the residual will drop down quickly in the first 2-10 steps. 25 steps were taken to get to a point where both our simulation and experiment have state covariance converging to a stable value, for presenting our results.

\begin{figure}[htbp]
\centering
\includegraphics[trim=0mm 5mm 15mm 5mm,clip=true,width=\linewidth]{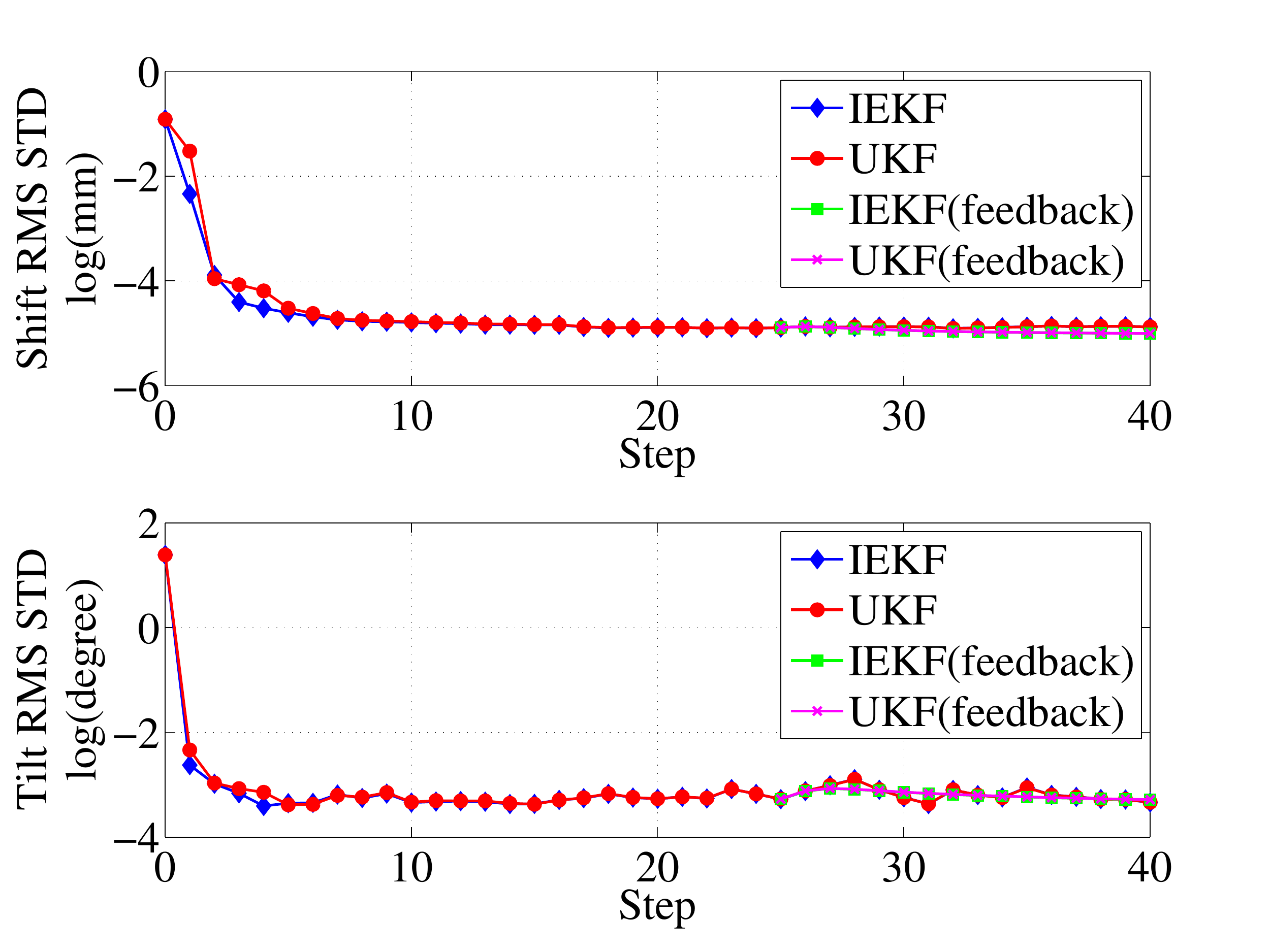}
\caption{RMS standard deviation (STD) of state estimation using IEFK and UKF in the simulation. Blue diamond line and red circle line represent IEKF and UKF estimation. Green square line and the magenta cross line represent the STD of state estimate with full state feedback after the 25 step in IEKF and UKF, respectively.}
\label{fig:kf_std}
\end{figure}

Figure \ref{fig:kf_std} shows the RMS standard deviation of state estimation using IEFK and UKF in the simulation. The standard deviation is square root of each diagonal elements of the state estimate covariance matrix $\mathbf{P}_k$ in IEKF and UKF. The lines with blue diamond and red circle markers represent IEKF and UKF estimation, respectively. In the initial estimation-only phase, the uncertainty in the state estimate can be seen to decrease rapidly in the first few steps. When we start feeding back the state estimate as a control input, the uncertainties become more stable as shown by the green square and magenta cross marked lines.

\subsection{Experiment Setup}
\begin{figure}[htbp]
\centering
\includegraphics[trim=40mm 40mm 40mm 30mm,clip=true,width=\linewidth]{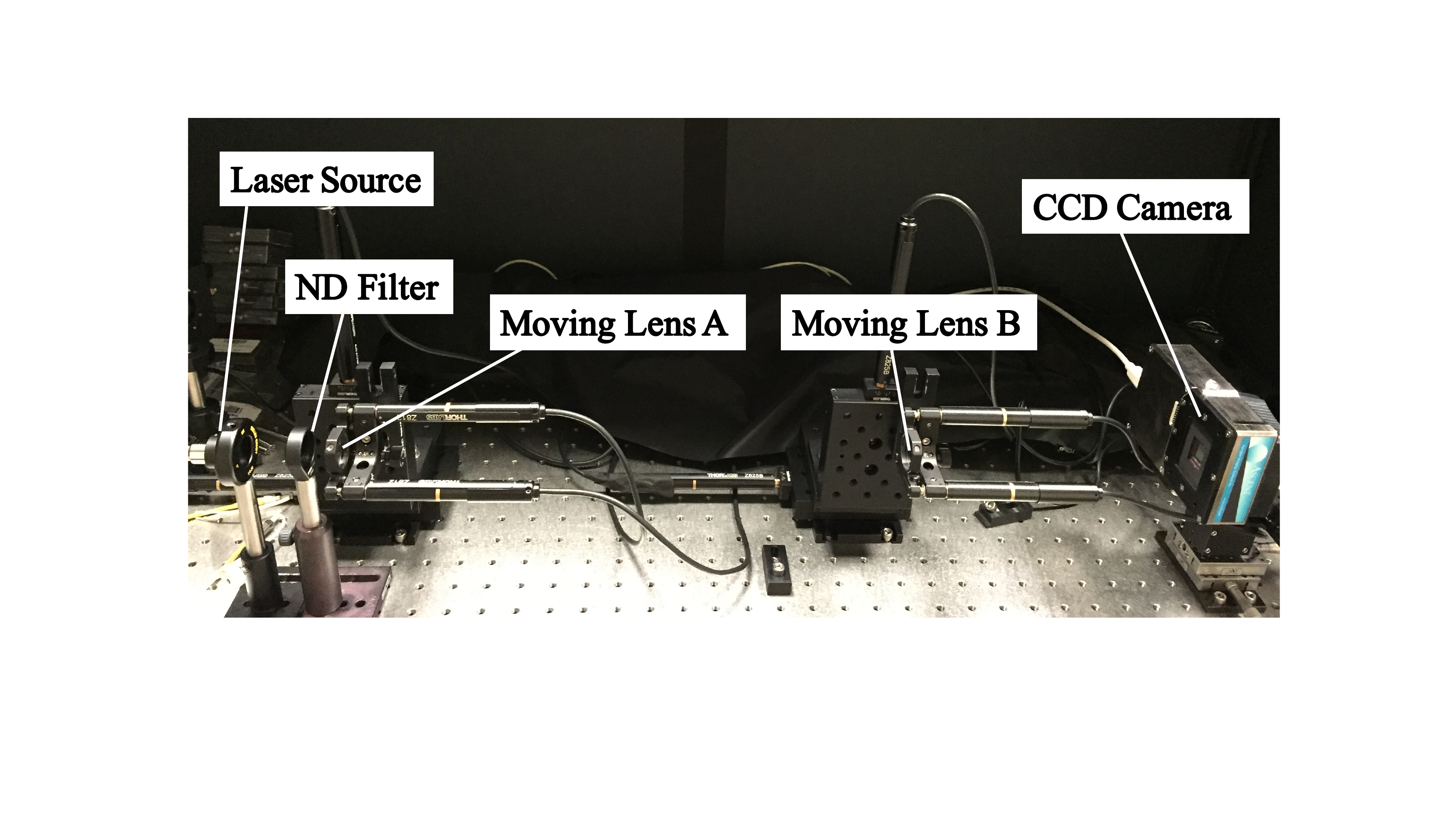}
\caption{Experiment setup of optical model shown in Figure \protect\ref{fig:two_lenses_model}. A collimated laser beam passes through a ND filter, two moving lenses A and B, and focuses on a CCD camera. The optical system after the ND filter is setup as the one simulated in ZEMAX.}
\label{fig:exp_setup}
\end{figure}

Figure \ref{fig:exp_setup} shows the experimental setup. The input 635 $nm$ laser beam is passed through a customized collimator to produce a collimated beam as in the simulation. A neutral density (ND) filter is installed after the collimator to reduce the power of the laser beam. The moving lens A is placed after the ND filter, and the moving lens B is placed 400 $mm$ after lens A. The CCD camera is 212 $mm$ away from lens B. Both lens A and B are mounted on motorized tip-tilt and translation stages. The stages and CCD camera are connected to a local computer which performs all data processing and can send actuation commands to the stages. Table \ref{tab:exp_spec} shows the detailed information of the components and devices in the experiment. The experimental results are presented in the next section.

\begin{table}
\caption{List of components and devices in the experiment}
\vspace{-1ex}
\label{tab:exp_spec} 
\begin{center}
\resizebox{1\columnwidth}{!}{
\begin{tabular}{l*{6}{c}r} Item & Model & Description \\ \hline 
Laser & Thorlabs MCLS1-635 & 635 $nm$ \\ 
Collimator & TC25FC-633 & Beam diameter ($1/e^2$): 4.67 $mm$ \\
ND filter & NE50B-A & OD: 5.0 \\
Lens A & LB1945-A & Focal length: 200 $mm$ \\
Lens B & LB1676-A & Focal length: 100 $mm$ \\
CCD camera & Apogee A694 & Pixel size: 4.54 $\mu m$, 16-bit\\
Translation stage & PT1-Z8 & Backlash < 8 $\mu m$\\
Tip-tilt stage & KS1-Z8 & Backlash < 8 $\mu m$ \\
\end{tabular}
}
\end{center}
\vspace{-4ex}
\end{table}

\section{Result}\label{sec:results}
In this section we present the image reconstruction in the experiment and the result of closed-loop control with Kalman filtering.

\subsection{Experiment Image Reconstruction}
\begin{figure}[htbp]
\centering
\includegraphics[trim=30mm 15mm 15mm 0mm,clip=true,width=\linewidth]{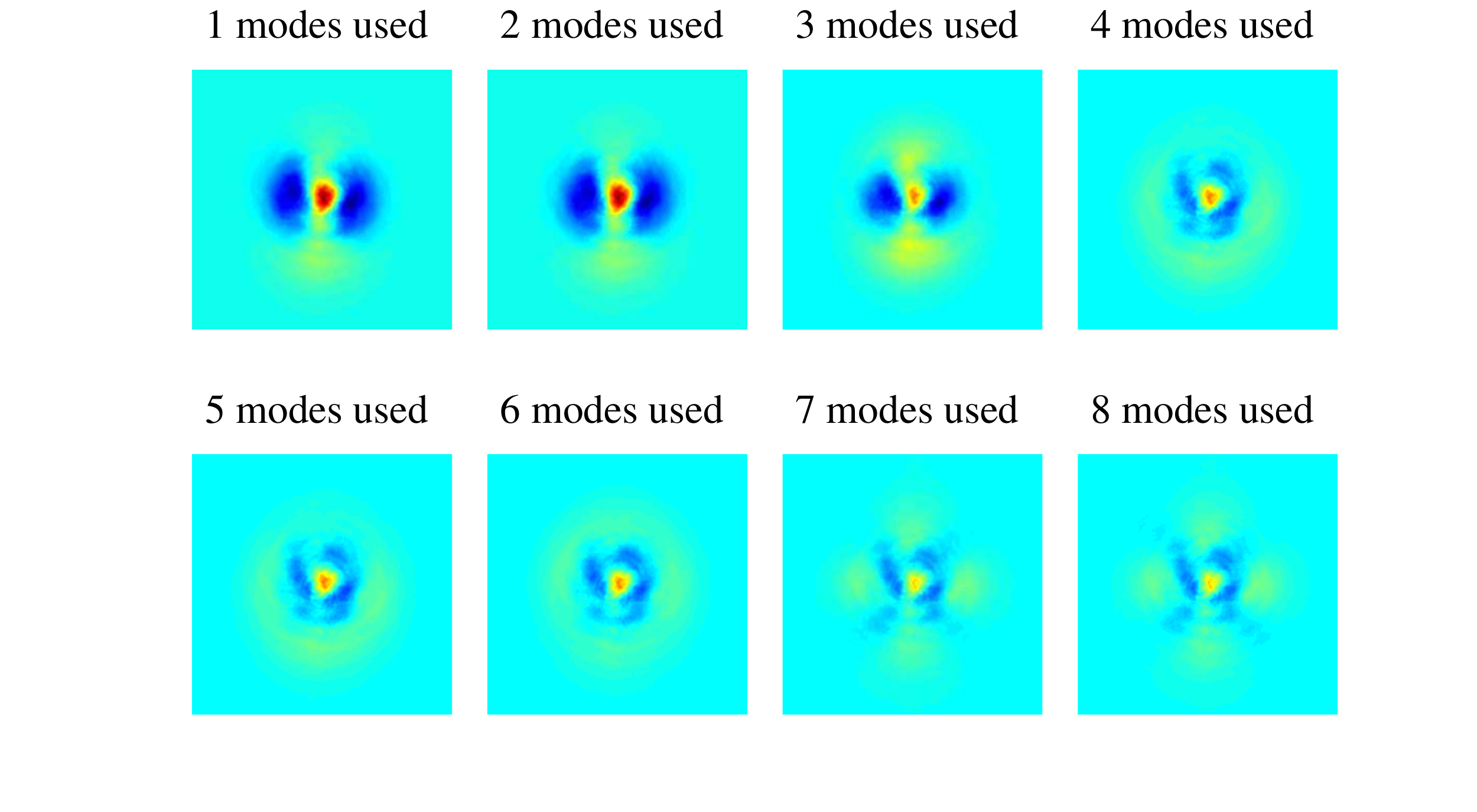}
\caption{Residual error of the reconstruction of a single images acquired with the experimental setup shown in Figure \protect\ref{fig:exp_setup}. The first eight KL modes derived in simulation are used. Images are plotted using the same intensity scale. As in simulation, the reconstruction error decreases gradually as the number of modes used increases.}
\label{fig:recon_exp}
\end{figure}

Figure \ref{fig:recon_exp} shows the residual error after reconstructing a single experimental image using the first eight KL modes derived in simulation. The corresponding sum of RMS pixel errors are shown as the red line with square markers in Figure \ref{fig:recon_rms_error}. The sum of RMS errors shown is the average over 500 images with random state inputs collected in the experiment. The reconstruction error in the experiment has the same trend as a function of KL mode number as in the simulation. The higher pixel error in the experiment is caused by the additional noise sources in the system, such as variations in the laser source, imperfections in the optics, and undamped vibrations in the lens stages.

\subsection{State Estimation Result}
\begin{figure}[htbp]
\centering
\includegraphics[trim=5mm 0mm 10mm 0mm,clip=true,width=\linewidth]{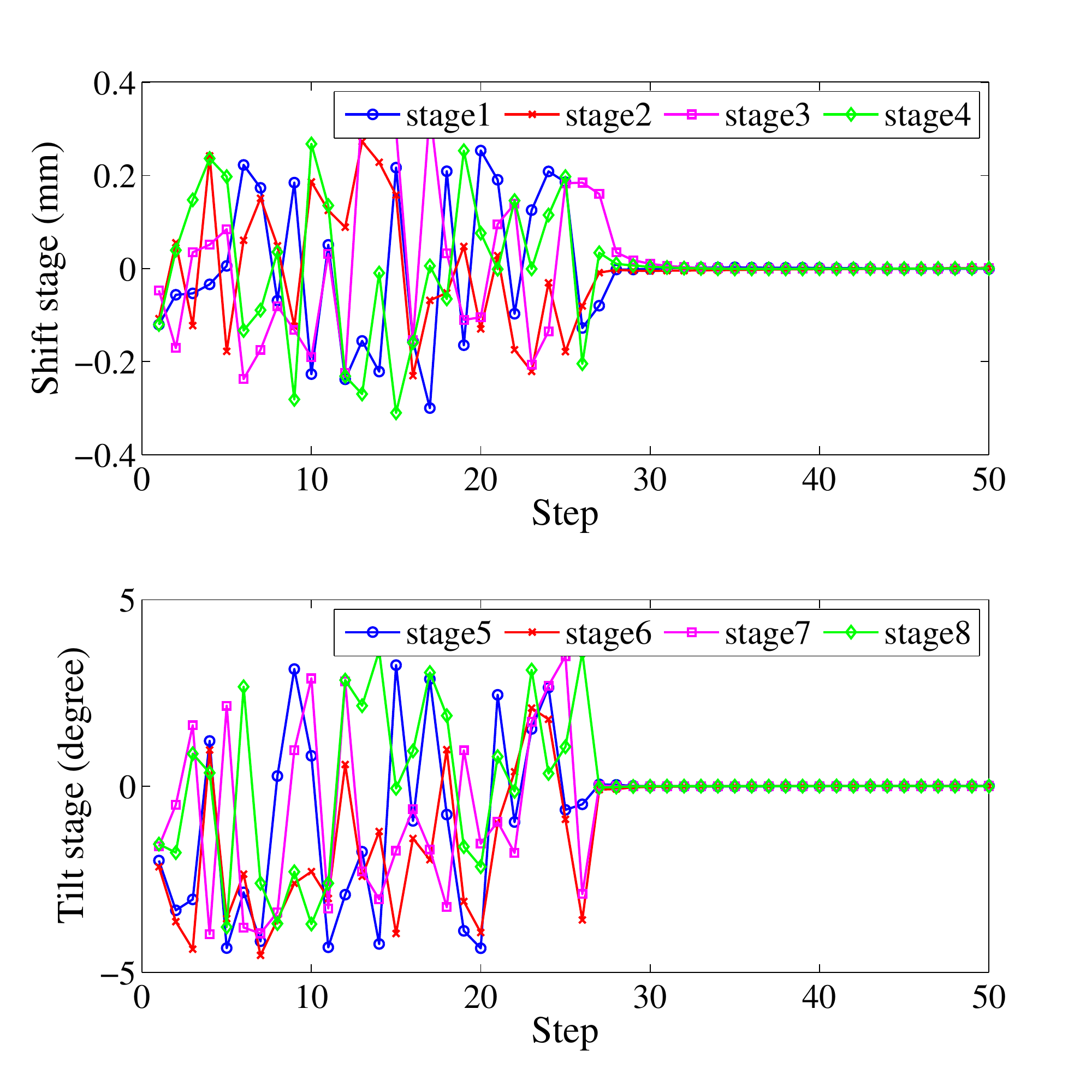}
\caption{Stage positions from step 1 to 50 using IEKF. Stages 1-4 correspond to the shift stages, and stages 5-8 are the tip and tilt stages driven by translation motors.}
\label{fig:stage_plot}
\end{figure}

In the experiment we give random input to the stages for the first 25 iteration steps, and then feed back the full state estimate as the control input $\mathbf{u}_k = -\hat{\mathbf{x}}_{k-1}$ from step 26 through 50. The state residual cannot be obtained in the experiment since the true state is unknown. Figure \ref{fig:stage_plot} shows the stage position as a function of iteration using the IEKF. Stages 1-4 correspond to state elements $\mathbf{x}_1$ to $\mathbf{x}_4$, and stages 5-8 are the tip and tilt stages driven by translation motors. We decrease the process noise covariance matrix to $\mathbf{Q}_k/4$ from step 26 on as the noise should be relatively small when the motor is moving in a small range. The stage positions converge to steady-state values as shown in Figure \ref{fig:stage_plot}. The UKF experiment has similar results to the IEKF in terms of stage position. 

\begin{figure}[htbp]
\centering
\includegraphics[trim=10mm 4mm 10mm 7mm,clip=true,width=\linewidth]{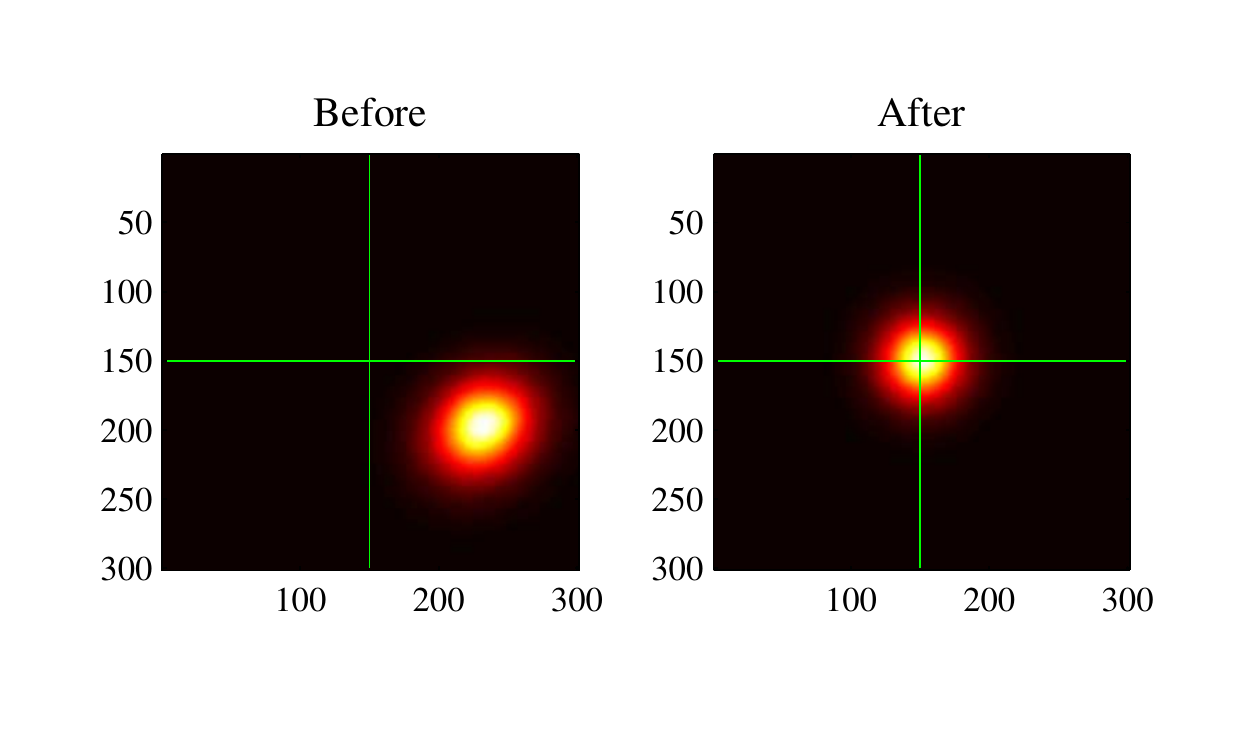}
\vspace{-5ex}
\caption{Experimental image before and after state feedback . The left image shows the $300 \times 300$ subframe before the correction, and the right image is the subframe after the correction. The intersection of the green lines represent the center of the camera.}
\label{fig:exp_img}
\end{figure}

Figure \ref{fig:exp_img} shows the initial and final experimental images, before and after closed-loop control. The left image shows the subframe before the state feedback, and the right images is the subframe after the state feedback converges. As expected, the image shifts to the center and becomes significantly more axisymmetric after state feedback converges.

\begin{figure}[htbp]
\centering
\includegraphics[trim=0mm 0mm 0mm 0mm,clip=true,width=\linewidth]{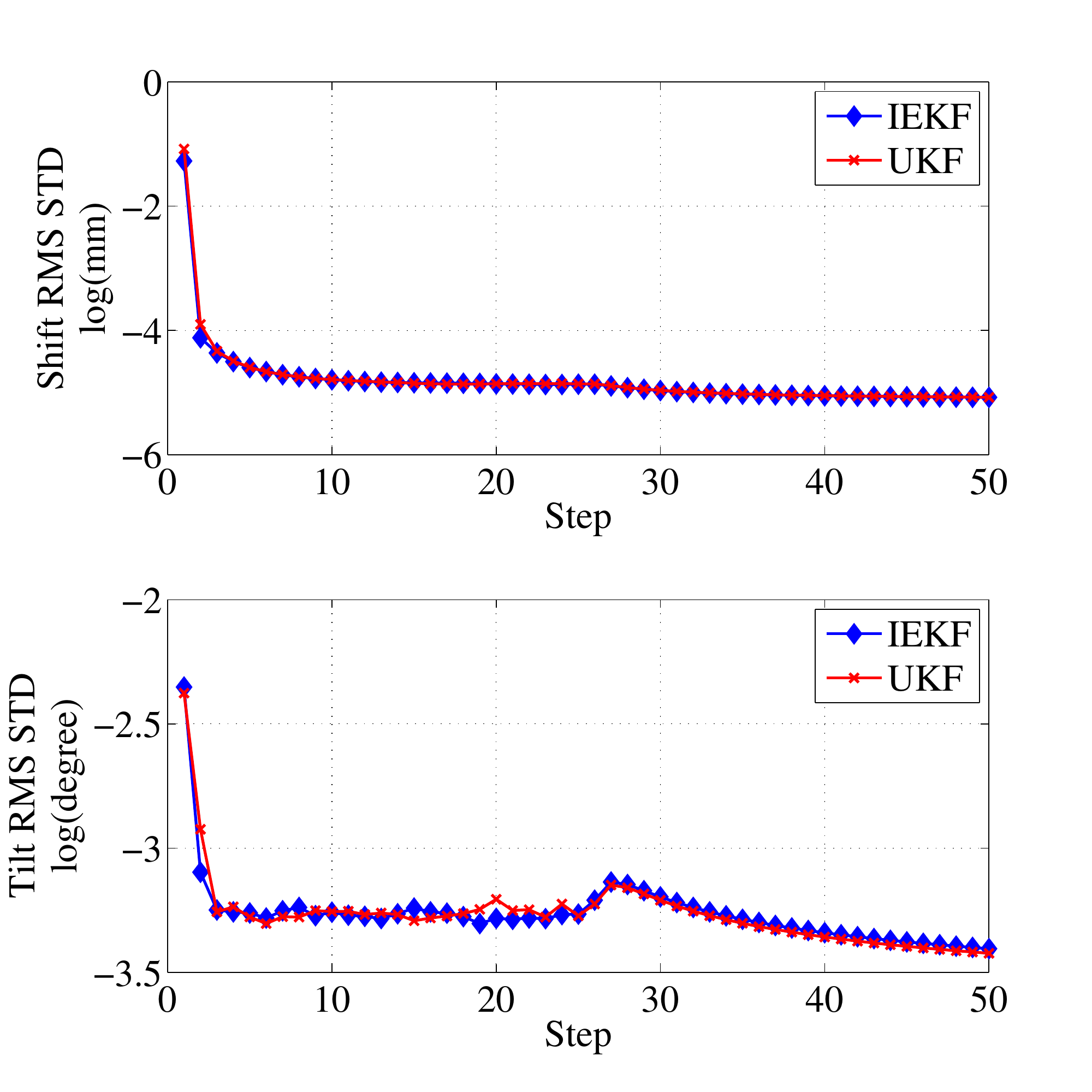}
\caption{RMS standard deviation of state estimate averaging over 20 experiments. Blue diamond line and red circle line represent IEKF and UKF respectively.}
\label{fig:exp_std}
\end{figure}

Figure \ref{fig:exp_std} shows the RMS standard deviation of state estimation averaging over 20 executions of the closed loop experiment. The lines with blue diamond and red circle markers represent the IEKF and UKF estimation, respectively. Similar to the simulation, the uncertainly drops down rapidly in the first few steps. The STD of tilt estimate increases slightly when we start feeding back the state estimates, but drops down gradually after a few steps.

Although the stage positions converge to stable values in a single test, the final stable values vary somewhat between different runs. When the states are close to zero it is likely to converge to a local minimum where the shifts, tip, and tilt of the two lenses compensate with each other. Table \ref{tab:exp} shows the standard deviation of stage convergence value in 20 tests. The variance of the final convergence values are of the same order. There is no evidence showing that UFK outperforms IEKF on our system, and vice versa. This indicates that the local linearization approximation in the IEKF is reasonable as UFK would capture the nonlinearity better than IEKF.

\begin{table}
\caption{STD of stage convergence value in 20 runs. Both IEKF and UKF are presented.}
\vspace{-1ex}
\label{tab:exp} 
\begin{center}
\resizebox{0.75\columnwidth}{!}{
\begin{tabular}{l*{6}{c}r}  & IEKF STD (mm) & UKF STD (mm) \\ \hline 
stage 1 & 0.0345 & 0.0335 \\ 
stage 2 & 0.0394 & 0.0454 \\
stage 3 & 0.0360 & 0.0344 \\
stage 4 & 0.0189 & 0.0232 \\
stage 5 & 0.1342 & 0.1163 \\
stage 6 & 0.1092 & 0.0845 \\
stage 7 & 0.1067 & 0.1469 \\
stage 8 & 0.0897 & 0.1071 \\
\end{tabular}
}
\end{center}
\vspace{-4ex}
\end{table}

\section{Conclusion and Future Work}\label{sec:discussion}
The automated alignment algorithm described here has two fundamental steps: 1. Image processing and reconstruction, and 2. State estimation and control. In the first step, the reconstruction error can be separated into the reconstruction bias and the detector noise. The bias is denoted as ($\bar{\mathbf{v}}_i - \bar{\mathbf{c}}_i$) in section \ref{sec:pca} and is neglected in the calculations for this step since it does not have a direct effect on the measurement error, and is considered as redundant information for correcting the system. In contrast, the noise in the image will be projected together with the image into the measurement $\mathbf{y}$ and becomes measurement noise ($\mathbf{R}_{meas}$) in the control and estimation step. 

There are two additional error sources in the control and estimation step which are the modelling error and the process noise ($\mathbf{Q}_k$). The modelling error includes the measurement function fitting error ($\mathbf{R}_{model}$) and inconsistencies between the simulated optical system and the experimental setup. The inconsistencies can include both errors in component placement as well as unmodeled effects such as thermal drifts, and these effects cannot be corrected with the current implementation of Kalman filtering. The process noise ($\mathbf{Q}_k$) includes the actuator repeatability, backlash, and stage hysteresis, etc. The overall experiment error is a combination of all of these error sources. The uncertainty of these factors results in the variation of the final stage position shown in Table \ref{tab:exp}. The less sensitive our lenses are to misalignment, the greater the variation of the stage positions will be. Different local minima are found in various iterations in the experiment. The variation of the pixel value between these iterations has an average standard deviation $\sim24$ in a $2^{16}$ dynamic range. The pixel with the highest STD ($\sim600$) occurs around the center of the each image.

Although $z$ axis misalignments of lenses A and B are ignored in the current model, the despace parameter ($z$ axis) is important in optical alignment. The component placement error might decrease if the $z$ axis movement is included in state variables. The movement corresponds to focus motion, and will occur mostly on axisymmetric KL modes such as Mode 3 (similar to defocus). A nonlinear function mapping from the despace parameter to the weight of KL modes can be learned. We expect the misalignment can be calibrated using an IEKF as long as the modeled nonlinear function has good performance using the local linearization approximation. If the learned measurement model function is highly nonlinear, we expect UKF will outperform IEKF, and should be used for state estimation. The misalignment of the focus motions will be included in future applications.

In simulation we have state residuals in multiple tests that show that the method is able to correct the misalignment to below a certain threshold. However, the method utilizes only an on-axis point source, and does not consider the alignment effects on optical aberrations across full FOV. Since many optical instruments use wide FOVs, a particularly important set of future tests will be to evaluate the effects of the variability of the final converged stage positions on off-axis point sources and images of extended sources.  To the extent that this variability represents the insensitivity of the whole imaging system to this level of misalignments, we would expect similar results for on-axis and off-axis sources.  However, if the different results actually represent truly different local minima where the lenses compensate for each others' misalignments in different ways, then we may expect different levels of distortion throughout the final image.

Our future work will also  focus on extending the concept of our self-aligning method into various systems and applications. One of the extensions is to apply the method to reflective optical systems. For any system model misalignments can be added into both reflective and refractive moving components and similar steps to the ones described here can be performed. Some specific examples of possible applications are aligning components in a two-mirror telescope, aligning optics with human eyes in virtual reality headsets, and adjusting subsystems in a two triplet systems. While aligning these optical systems, moving components in optical systems in practice are often restricted by the performance of the actuators. The actuators are limited by their repeatability and backlash. Further investigation on the limitation of actuators affecting the system to meet the performance specification is required.

\bigskip

\bibliography{ao_v3}

\ifthenelse{\equal{\journalref}{ol}}{%
\clearpage
\bibliographyfullrefs{articlebib}
}{}


\appendix
\section*{Appendix A: Measurement function}
\begin{table}[h!]                                                                            
\centering  
\scalebox{0.55}{
\begin{tabular}{|c|c|c|c|c|c|c|c|}                                                              
\hline                                                                                          
 & $y_1$ & $y_2$ & $y_3$ & $y_4$ & $y_5$ & $y_6$ & $y_7$ \\                                                          
\hline                                                                                         
$x_1$ & -2.46E-04 & 8.86E-04 & -1.04E-04 & 1.90E+00 & 1.93E-03 & -1.22E+00 & -2.01E-04 \\      
\hline                                                                                         
$x_2$ & -2.30E-04 & 1.17E-04 & -5.00E-05 & -9.01E-04 & -1.90E+00 & -1.50E-04 & -1.22E+00 \\    
\hline                                                                                         
$x_3$ & -5.65E-04 & -7.05E-04 & -5.48E-04 & -8.57E-01 & 4.81E-04 & 2.14E+00 & 5.49E-05 \\      
\hline                                                                                         
$x_4$ & 1.15E-04 & 6.05E-04 & 1.34E-04 & -3.28E-04 & 8.59E-01 & 2.75E-05 & 2.14E+00 \\         
\hline                                                                                         
$x_5$ & 5.13E-05 & 7.66E-05 & -8.70E-05 & 6.86E-05 & -2.51E-01 & 1.13E-06 & 8.78E-03 \\        
\hline                                                                                         
$x_6$ & 2.52E-06 & -1.01E-04 & -4.13E-05 & -2.51E-01 & 4.60E-05 & -8.79E-03 & -4.72E-07 \\     
\hline                                                                                         
$x_7$ & 1.05E-05 & -6.08E-05 & 4.11E-05 & -2.95E-05 & 2.29E-02 & -7.12E-06 & 1.67E-03 \\       
\hline                                                                                         
$x_8$ & 1.57E-05 & 9.40E-05 & 1.05E-04 & 2.35E-02 & 3.63E-05 & -1.64E-03 & -4.74E-06 \\        
\hline                                                                                         
$x_1^2$ & -1.08E-03 & -3.19E-01 & -2.48E-01 & -2.44E-04 & 7.10E-03 & 1.72E-03 & -8.17E-04 \\   
\hline                                                                                         
$x_2^2$ & -2.90E-03 & -3.23E-01 & 2.52E-01 & -1.48E-03 & 3.74E-03 & -1.61E-04 & -3.53E-05 \\   
\hline                                                                                         
$x_3^2$ & 7.23E-04 & -8.42E-02 & -5.87E-02 & 2.45E-03 & 1.48E-04 & 9.00E-05 & 4.85E-05 \\      
\hline                                                                                         
$x_4^2$ & -1.12E-03 & -8.26E-02 & 6.23E-02 & 1.97E-03 & -2.22E-03 & -4.46E-05 & 2.20E-04 \\    
\hline                                                                                         
$x_5^2$ & -3.38E-06 & -4.87E-02 & 4.71E-02 & -1.03E-04 & 1.87E-05 & 2.16E-05 & 1.02E-06 \\     
\hline                                                                                         
$x_6^2$ & 2.43E-06 & -4.91E-02 & -4.66E-02 & -4.34E-05 & -2.03E-05 & 6.66E-05 & 4.71E-06 \\    
\hline                                                                                         
$x_7^2$ & 1.87E-05 & -8.41E-02 & 8.23E-02 & -4.59E-05 & 3.87E-05 & 5.33E-05 & 3.89E-07 \\      
\hline                                                                                         
$x_8^2$ & 1.77E-05 & -8.47E-02 & -8.13E-02 & 4.53E-04 & 1.49E-05 & 2.24E-04 & -1.06E-06 \\     
\hline                                                                                         
$x_1$$x_2$ & 2.87E-01 & 2.19E-03 & -3.14E-03 & -2.37E-03 & 9.00E-04 & -1.11E-03 & 7.04E-04 \\  
\hline                                                                                         
$x_1$$x_3$ & -3.28E-03 & 3.17E-01 & 2.37E-01 & -1.43E-03 & -1.78E-03 & -1.56E-03 & 5.77E-04 \\ 
\hline                                                                                         
$x_1$$x_4$ & -1.40E-01 & 7.92E-04 & 1.22E-03 & -2.84E-03 & 6.16E-03 & -1.50E-04 & -2.14E-04 \\ 
\hline                                                                                         
$x_1$$x_5$ & 1.85E-02 & -9.71E-05 & -1.05E-04 & 4.23E-04 & -2.48E-04 & -8.86E-06 & 3.41E-05 \\ 
\hline                                                                                         
$x_1$$x_6$ & -1.06E-04 & 3.85E-02 & 3.04E-02 & -4.59E-04 & 4.88E-04 & 1.12E-06 & -8.89E-06 \\  
\hline                                                                                         
$x_1$$x_7$ & -2.92E-02 & 1.60E-04 & -8.62E-05 & 1.24E-04 & 6.98E-04 & -2.89E-05 & -9.43E-05 \\ 
\hline                                                                                         
$x_1$$x_8$ & 9.20E-05 & -4.60E-02 & -4.43E-02 & 1.37E-03 & 3.19E-04 & 2.93E-04 & -1.25E-05 \\  
\hline                                                                                         
$x_2$$x_3$ & -1.41E-01 & 5.63E-04 & 3.16E-04 & 1.64E-03 & 1.36E-03 & 2.08E-04 & -4.62E-04 \\   
\hline                                                                                         
$x_2$$x_4$ & -1.53E-04 & 3.18E-01 & -2.46E-01 & -2.85E-04 & 1.05E-03 & -1.07E-04 & 9.03E-05 \\ 
\hline                                                                                         
$x_2$$x_5$ & -2.03E-04 & -3.89E-02 & 3.12E-02 & 3.62E-04 & 2.00E-05 & 2.87E-05 & -9.66E-05 \\  
\hline                                                                                         
$x_2$$x_6$ & -1.86E-02 & -1.21E-04 & 3.16E-05 & 2.55E-04 & -1.34E-04 & 1.39E-05 & 3.65E-05 \\  
\hline                                                                                         
$x_2$$x_7$ & 1.68E-04 & 4.57E-02 & -4.54E-02 & -3.71E-04 & -2.50E-04 & -5.25E-05 & -3.12E-05 \\
\hline                                                                                         
$x_2$$x_8$ & 2.92E-02 & 4.68E-04 & 1.92E-04 & 8.76E-05 & -2.62E-04 & 2.17E-05 & 7.26E-05 \\    
\hline                                                                                         
$x_3$$x_4$ & 7.26E-02 & 1.80E-03 & 7.09E-04 & 5.48E-04 & 1.17E-03 & 1.24E-04 & 3.78E-04 \\     
\hline                                                                                         
$x_3$$x_5$ & 6.04E-04 & 4.38E-04 & -1.39E-04 & 3.10E-04 & -3.65E-05 & 6.13E-05 & -1.91E-05 \\  
\hline                                                                                         
$x_3$$x_6$ & 1.14E-04 & 1.98E-03 & 1.47E-03 & 4.33E-05 & 7.07E-05 & -3.45E-06 & 1.28E-05 \\    
\hline                                                                                         
$x_3$$x_7$ & 2.69E-02 & -1.78E-04 & -4.17E-06 & 2.42E-04 & -6.55E-04 & 7.63E-05 & 1.05E-04 \\  
\hline                                                                                         
$x_3$$x_8$ & -1.19E-04 & 4.58E-02 & 4.45E-02 & -1.04E-03 & -2.08E-04 & -2.11E-04 & -6.04E-05 \\
\hline                                                                                         
$x_4$$x_5$ & -1.63E-04 & -1.82E-03 & 1.52E-03 & -4.68E-04 & 2.18E-04 & -4.55E-05 & -4.26E-06 \\
\hline                                                                                         
$x_4$$x_6$ & -8.67E-04 & 3.02E-05 & -3.04E-04 & 2.54E-04 & -1.11E-04 & 8.12E-06 & -1.17E-05 \\ 
\hline                                                                                         
$x_4$$x_7$ & -8.33E-06 & -4.60E-02 & 4.53E-02 & 1.46E-04 & 1.79E-05 & 8.46E-05 & 2.51E-05 \\   
\hline                                                                                         
$x_4$$x_8$ & -2.73E-02 & 2.77E-04 & 1.70E-04 & 5.70E-05 & 4.93E-04 & -3.83E-05 & -7.77E-05 \\  
\hline                                                                                         
$x_5$$x_6$ & -5.32E-02 & 9.31E-06 & 2.66E-05 & 6.23E-06 & 1.12E-04 & -3.86E-06 & -4.62E-05 \\  
\hline                                                                                         
$x_5$$x_7$ & -8.74E-06 & -3.95E-04 & 3.92E-04 & -3.45E-05 & -3.64E-05 & -4.61E-06 & 4.93E-06 \\
\hline                                                                                         
$x_5$$x_8$ & -2.02E-04 & -2.29E-07 & 3.28E-06 & -2.43E-05 & -2.50E-05 & -1.23E-06 & 3.94E-06 \\
\hline                                                                                         
$x_6$$x_7$ & -1.97E-04 & -3.10E-05 & -1.32E-05 & -2.72E-05 & 2.75E-05 & -2.46E-06 & 2.04E-06 \\
\hline                                                                                         
$x_6$$x_8$ & -2.43E-05 & -3.89E-04 & -3.60E-04 & 6.82E-06 & -1.93E-05 & 2.21E-06 & 3.93E-06 \\ 
\hline                                                                                         
$x_7$$x_8$ & -9.21E-02 & 1.15E-05 & -7.16E-07 & 2.71E-05 & 5.88E-04 & 2.66E-06 & -1.68E-04 \\  
\hline                                                                                         
constant & -4.63E-05 & 2.68E+00 & -1.37E-02 & -1.60E-03 & -3.08E-04 & -1.96E-03 & -2.83E-05 \\ 
\hline                                                                                         
\end{tabular}  
}
\caption{Coefficients of nonlinear measurement function $\mathbf{h}(\mathbf{x})$. $x_1$ to $x_8$ represent the elements in state vector $\mathbf{x}$. $y_1$ to $y_7$ are the seven measurements.}                                                                      
\label{table:h}                                                                     
\end{table}

\end{document}